\documentclass[aps,prb,twocolumn,groupedaddress]{revtex4}
\usepackage{graphics}

\begin{document}

\title{The time dependence of muon spin relaxation spectra and spin correlation functions}
\author{C. A. Steer}
\address{ Department of Physics, University of Warwick, Coventry  CV4 7AL, United Kingdom.}
%% \author{S. J. Blundell}
%% \address{ Department of Physics, University of Oxford, Oxford  OX1  7AL.}

\begin{abstract}
The existing theory of the microscopic interpretation of the dynamical contribution to zero-field muon depolarization spectra in a longitudinal geometry is developed. The predicted relaxation of the muon depolarization is calculated from two forms of the spin correlation function. First, when the spin correlation function has an exponential form with a single wave vector dependent relaxation rate is considered, it is shown that this form of the spin correlation function reproduces the slow and fast fluctuation limits of stochastic spin theory regardless of the choice of microscopic spin model. Second, if the spin correlation function is a homogeneous scaling function (such as a power-law decay with time), as suggested by the mode-coupling theory of spin dynamics, this results in a stretched exponential relaxation of the muon spectra. For simple spin diffusion, the muon spectra are shown to be relax with a root-exponential form. 
\end{abstract}
\maketitle
\section{Introduction}
The study of spin dynamics is fundamental to the classification of magnetic materials. One tool that has been used extensively is neutron scattering which has provided experimentalists with detailed knowledge of the both static and dynamic properties of magnetic systems. However neutron scattering is inappropriate for samples with a large absorption cross-section, or which are only available in small quantities, or when the magnetic moment is small. 

Muon spin relaxation spectroscopy ($\mu$SR) is an complementary probe of the spin dynamics in magnetic materials, and can be used for materials when neutron spectroscopy is not suitable\cite{spinreview}. The muon is implanted into a magnetic lattice and is sensitive to the spin fluctuations through the fluctuations of the local magnetic field. The motion of the muon spin is sensed when it decays into a positron and other products. The positron is emitted preferentially in the direction of the muon spin at the time of its decay. This can be studied in longitudinal muon relaxation spectroscopy where the emitted positrons are counted in two banks of detectors, one in front and one behind the sample position. The quantity of experimental interest is the time-dependent muon asymmetry which measures the time dependence of the motion of the muon depolarization. This is given by
\begin{equation}
P_{Z} (t) = \frac{N_{F} (t) - \alpha N_{B} (t)}{N_{F} (t) + \alpha N_{B} (t)},
\end{equation}
where $N_{F} (t)$ is the number of positron counts in front bank of detectors, $N_{B} (t)$ is the number of positron counts in the rear bank of detectors and $\alpha$ is an experimentally determined constant. 

The longitudinal asymmetry function, $P_{Z} (t)$, can have static and dynamic contributions. In the case of completely static local fields, the muon senses an average of the local field given by 
\begin{equation}
  P_{Z} (t) = \int \rho({\bf B}_{\mu}) \left\{ \cos^{2} \theta + \sin^{2}( \theta )\cos (\gamma_{\mu} B_{\mu} t) \right\} d{\bf B}_{\mu},
\end{equation}
where $\theta$ is the angle between the local field (given by the vector ${\bf B}_{\mu}$) and the initial muon spin, $\rho ({\bf B}_{\mu})$ is the magnetic field distribution function and $\gamma_{\mu}$ is the muon's gyromagnetic ratio (2$\pi \times$ 13.55 kHz /G).  The choice of field distributions involves some motivation from the knowledge of the magnetic system. For instance for a dilute spin glass the field distribution is likely to be a Gaussian distribution, which gives rise to a Kubo-Toyabe relaxation function \cite{kt}. One other relaxation function of note, as it occurs often, is the stretched exponential\cite{steer} ($P_{Z} (t) \propto \exp\left\{-(\lambda t)^{\kappa}\right\}$). The form where $\kappa = \frac{1}{2}$ is often seen in the low temperature properties of various types of magnetic materials \cite{blundell,jestadt,bewley} and is attributed to the Uemura distribution of local fields \cite{uemura}. 

This paper is concerned with the amount of information that can be obtained about {\it dynamic} spin fluctuations from muon spectra. Previous theoretical studies of the dynamic contribution to the shape of the muon asymmetry have predicted a simple exponential form, as they have been studied in the fast fluctuation limit (when the spins fluctuate over times much shorter than the muon experimental time window)\cite{yaouanc}. These studies established a direct relationship of the spin correlation function and the muon asymmetry. However, it has been known for some time that the spin correlation function may exhibit long-time power-law tails\cite{sw1,hubbardjap} that may be significant in the muon time window. 

It is possible to identify two regimes for spin correlation function, the high temperature paramagnetic regime, where the spin correlation function exponentially decays\cite{blumehubb} and the critical regime where it is a homogeneous function\cite{note} of the scaling variable $q^{-\theta}t$ (where $q$ is the wave vector, $t$ is time and $\theta$ is an exponent)\cite{sw1}. In this paper the dynamic contribution to the time-dependence of the muon asymmetry is derived, without using the fast fluctuation limit, for these two common forms of the spin correlation function. 

In the next section the theory describing how spin fluctuations affect muon spectra are briefly reviewed. A method of Laplace transforming the muon polarization function is used and this is applied to the case when the spin fluctuations are decaying exponentially with time. Mode-coupling theory suggests that the spin correlation function should be a homogeneous function of the product $q^{\theta}t$ (where $q$ is the wave vector and $t$ is time) so the effect on the depolarization spectra of this form is studied. It is shown that this can result in a stretched exponential relaxation and, in the case of simple spin diffusion, root exponential behavior. The exponent ($\kappa$) is dependent on the model of spin dynamics chosen. 

\section{A brief review of existing theory}

The muon depolarization shape has been calculated using perturbation theory. A quantum calculation \cite{dy} and a semi-classical calculation \cite{keren1} lead to
\begin{equation}
P_{z} (t) = \exp \left\{ - \Psi_{z} (t)\right\},
\end{equation}
where the coordinate axes are such that the $z$-axis is aligned with the initial muon spin direction. The relaxation function $\Psi_{z}$ is given by
\begin{equation}
\Psi_{z} (t) = \gamma_{\mu}^{2} \int_{0}^{t} d\tau \left( t - \tau \right) \left[ \Phi_{xx} (\tau ) + \Phi_{yy} (\tau)\right],
\label{eq1}
\end{equation}
where $\Phi_{\alpha \alpha} (t) = \frac{1}{2} \langle \delta B_{\alpha}(t) \delta B_{\alpha} +  \delta B_{\alpha} \delta B_{\alpha} (t) \rangle$ is the field fluctuation correlation function for component $\alpha$ of the local magnetic field $B_{\alpha}$. This is a result of perturbation theory so it is valid when the interaction term of the muon and local field is much smaller than the other terms in the Hamiltonian. 

The dynamical behavior of the local field has been analysed before in the fast fluctuation limit when the field correlation functions are significant over a much shorter timescale than the muon experimental time window \cite{yaouanc}. This means that one can say that equation \ref{eq1} is proportional to time and the relaxation rate ($\lambda_{z}$), in this fast fluctuation limit, is given by\cite{yaouanc}
\begin{equation}
\lambda_{z} = \frac{\pi {\cal D}}{V} \int \sum_{\beta \gamma} {\cal A}^{\beta \gamma} ({\bf q}) \Lambda^{\beta \gamma} ({\bf q}, \omega_{\mu} = 0) \frac{d^{3} {\bf q}}{(2\pi )^{3}},
\label{lamz}
\end{equation}
where ${\cal D} = (\mu_{0} \gamma_{\mu} g \mu_{B} / 4 \pi)^{2} $, ${\cal A}$ is a tensor describing how the spins couple to the local magnetic field, $\Lambda ({\bf q}, \omega_{\mu} = 0)$ is the q-dependent spin correlation function approximately at zero-energy because of the low energy nature of the muon probe, $g$ is the usual Land\'{e} g-factor, $\mu_{B}$ is the Bohr magneton, $\gamma_{\mu}$ is the muon gyromagnetic ratio and $V$ is the volume of the sample. Also the wave vector integration in equation \ref{lamz} is over the first Brillouin zone. 

The time-dependence of the muon depolarization function, $P_{z} (t)$, is not necessarily just a simple exponential and this paper is concerned with the predictions of the shape of the muon depolarization equation that can be made from equation \ref{eq1}. Laplace transforming equation \ref{eq1} gives 
\begin{equation}
\Psi_{z} (s) = \gamma_{\mu}^{2} \frac{1}{s^{2}} \left\{\Phi_{xx} (s) + \Phi_{yy} (s)\right\}. 
\label{psiz}
\end{equation}
where $s$ is the Laplace conjugate variable to $t$. The Laplace-transformed field correlation functions are then given by
\begin{eqnarray}
\left\{\Phi_{xx} (s) + \Phi_{yy} (s)\right\} = \left\{ \frac{\mu_{0} g \mu_{B} }{4 \pi }  \right\}^{2} \frac{1}{V} && \nonumber \\ \times \int_{BZ} \sum_{\beta, \gamma = x,y,z} {\cal A}^{\beta \gamma} \Lambda^{\beta \gamma} (q, s) \frac{d^{D}q}{(2\pi )^{D}}, &&
\label{corr}
\end{eqnarray}
where ${\cal A}^{\beta \gamma} = G^{x \beta} (q) G^{x \gamma} (-q) + G^{y \beta} (q) G^{y \gamma} (-q)$ (where $\beta = x,y,z$ and $\gamma=x,y,z$) and $D$ is the dimensionality. This function $G^{\alpha \beta}$ is the Fourier transform of the dipole-dipole and hyperfine interactions between muon and lattice spins. Near $q \simeq 0$ this is given by\cite{yaouanc}
\begin{equation}
G^{\alpha \beta} (q) = - 4 \pi \left\{ P_{L}^{\alpha \beta} (q) - C^{\alpha \beta} (0) - \frac{r_{\mu} H \delta^{\alpha \beta}}{4\pi} \right\}
\label{G}
\end{equation}
where $H$ is the Hyperfine field at the muon site, $r_{\mu}$ is the number of nearest neighbours to the muon, $C^{\alpha \beta} (0)$ is a symmetric tensor\cite{yaouanc} and $ P_{L}^{\alpha \beta} (q) = q^{\alpha} q^{\beta} / q^{2}$. In this long wavelength limit the coupling tensor ${\cal A}^{\beta \gamma}$ is solely dependent on the angular variables of the integral. Using spherical coordinates in the integral gives 
\begin{eqnarray}
\left\{\Phi_{xx} (s) + \Phi_{yy} (s)\right\} &=& \left\{ \frac{\mu_{0}}{4 \pi } g \mu_{B} \right\}^{2} \frac{1}{V} \nonumber \\ \times \int_{0}^{q_{BZ}} \sum_{\beta, \gamma = x,y,z} && {\cal B}^{\beta \gamma}  q^{D -1} \Lambda^{\beta \gamma} (q, s) \frac{dq}{(2\pi )^{D}}
\label{corr2}
\end{eqnarray}
where the angular part of the integration is contained within the co-efficient ${\cal B}^{\beta \gamma}$, which is given by
\begin{equation}
{\cal B}^{\beta \gamma}  = \int_{\Omega_{q}} {\cal A}^{\beta \gamma} d\Omega_{q}.
\end{equation}

\section{Exponentially decaying spin correlations}

At high temperatures the spin correlation function for a simple ferromagnet has been shown to be an exponential function of time\cite{blumehubb,hubbardjap}. Therefore, the spin correlation function is given by
\begin{equation}
  \Lambda^{\beta \gamma} (q, t) = \frac{N k_{B} T}{g^{2} \mu_{0} \mu_{B}^{2} } \chi^{\beta \gamma} (q) e^{- \Gamma^{\beta \gamma} (q) t}. 
\end{equation}
where $N$ is the number of spins, $T$ is the temperature and  $\Gamma^{\beta \gamma} ({\bf q})$ is the $q$-dependent relaxation rate. Laplace transforming and substituting this into equations \ref{corr2} and \ref{psiz} gives
\begin{eqnarray}
\Psi_{z} (s) &=&  \frac{\gamma_{\mu}^{2}  \mu_{0} k_{B} T }{2^{4+D} \pi^{2+D} v } \sum_{\beta, \gamma = x,y,z} {\cal B}^{\beta \gamma} \nonumber \\ &\times& \int_{0}^{q_{BZ}} \frac{q^{D-1}\chi^{\beta \gamma} (q)  }{s^{2} \left[ s + \Gamma^{\beta \gamma} (q) \right]} dq. 
\label{psiz2}
\end{eqnarray}
where $v$ is the volume per spin. Inverse Laplace transforming this gives
\begin{widetext}
\begin{equation}
\Psi_{z} (t) = \Psi_{z}^{(0)}  \sum_{\beta, \gamma = x,y,z} {\cal B}^{\beta \gamma} \int_{0}^{q_{BZ}}        \frac{ q^{D-1} \chi^{\beta \gamma} (q)}{[\Gamma^{\beta \gamma} (q) ]^{2}} \left\{ e^{ - \Gamma^{\beta \gamma} (q) t} - 1 + \Gamma^{\beta \gamma} (q) t \right\} dq.
\label{psiz3}
\end{equation}
\end{widetext}
where $\Psi_{z}^{(0)} =(\gamma_{\mu}^{2}  \mu_{0} k_{B} T) /( 2^{4+D} \pi^{2+D} v ) $. The derived muon relaxation function is a q-integrated Abragam function\cite{abragam}. When the relaxation rate is independent of wave vector, this result also reduces to the result of Keren\cite{kerenDKT} in zero-field. The present extension however derives this result from a microscopic viewpoint as suggested by Yaouanc and Dalmas de Ro{\'e}tier\cite{commentddy}.

Stochastic spin dynamical theory \cite{hayano} predicts that in the fast and slow fluctuation limits the muon depolarization spectra are a simple exponential and Gaussian respectively. It is possible to consider the fast and slow fluctuation limits of equation \ref{psiz3}. In the fast fluctuation limit, which is when $\Gamma^{\beta \gamma} (q) t \gg 1$, the expression in the curly brackets of equation \ref{psiz3} tends to $\Gamma (q) t$ so the relaxation function reduces to fast fluctuation limit expression\cite{yaouanc}, and is given by
\begin{equation}
  \Psi_{z} (t) \simeq \Psi_{z}^{(0)} \hspace{0.1in} t \sum_{\beta, \gamma} {\cal B}^{\beta \gamma} \int_{0}^{q_{BZ}} q^{D-1} \frac{\chi^{\beta \gamma} (q)}{\Gamma^{\beta \gamma}(q)}  dq.
\end{equation}
In this limit the depolarization spectra are a simple exponential. In the slow fluctuation limit ($\Gamma^{\beta \gamma} (q) t \ll 1$)  the expression in the curly brackets of equation \ref{psiz3} tends to $\frac{1}{2}(\Gamma (q) t)^2$ so the relaxation function becomes
\begin{equation}
  \Psi_{z} (t) \simeq \frac{1}{2} \Psi_{z}^{(0)} \hspace{0.01in} t^{2} \sum_{\beta, \gamma} {\cal B}^{\beta \gamma} \left\{ \int_{0}^{q_{BZ}} \chi^{\beta \gamma} (q) q^{D-1}  dq\right\}.
\label{bollox}
\end{equation}
This analysis suggests that the spectra can be analysed continuously using equation \ref{psiz3} as the fluctuations slow down and pass through the muon experimental time window. The transition between these two limits will be dependent on the spin model chosen. 

In order to gain some insight the contribution to the muon relaxation spectra from one component can be calculated. The depolarization function can be written as
\begin{equation}
P_{Z} (t) = P_{Z} (0) \prod_{\beta \gamma} \exp \left\{ - \Psi^{\beta \gamma} (t) \right\}.
\label{prod}
\end{equation}
Here $\Psi^{\beta \gamma} (t)$ is given by equation \ref{psiz3} for one value of $\beta \gamma$. Then we use the Ornstein-Zernike form of the susceptibility $\chi^{\beta \gamma} (q) = \chi^{\beta \gamma} (0) \xi^{2} / (1 + (q\xi)^{2})$, where $\xi$ is the magnetic correlation length. This form for the $q$-dependent susceptibility has been used in the derivation of equation \ref{lamz}, and in previous studies of spin dynamics with $\mu$SR\cite{yaouanc}.

Also for three dimensional paramagnets at high temperatures, Blume and Hubbard\cite{blumehubb} found that the spin correlation function, at long times, has a fluctuation rate given by $\Gamma^{\beta \gamma} (q) = \Gamma q^{2}$ (where $\Gamma$ is a fundamental spin fluctuation rate). The prefactors ($\Psi_{z}^{(0)} {\cal B}^{\beta \gamma} \chi^{\beta \gamma} (0)$) are set to 1 as we are only interested in the shape change as the system passes through a spin-freezing transition.  

\begin{figure}
\centering
  \resizebox{80mm}{!}{\includegraphics{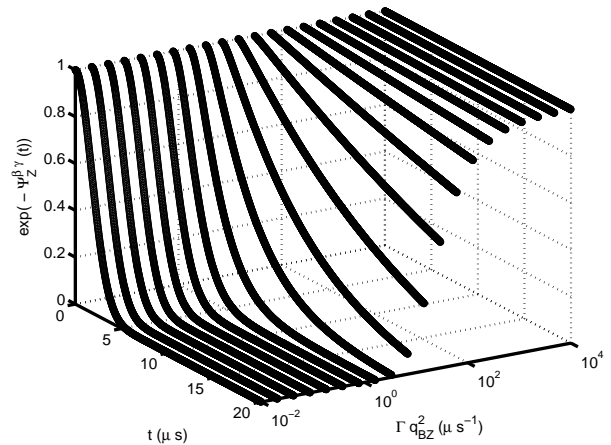}}
  \caption{The contribution to the muon relaxation function from one spin component for a three-dimensional paramagnet obeying $\Gamma^{\beta \gamma} (q) = \Gamma q^{2}$. The transition from a Gaussian to simple exponential relaxation function can be seen as the scaled spin fluctuation rate ($\Gamma q^{2}_{BZ}$) increases. This is plotted for a single value of the magnetic correlation length $q \xi = 0.1$.}
  \label{bigkahuna}

\end{figure}

 Using this and numerically integrating over the Brillouin zone in equation \ref{psiz3} we can find the normalised relaxation function $\exp \left\{ - \Psi^{\beta \gamma} (t) \right\}$ which is plotted in figure \ref{bigkahuna} for one value of the correlation length ($q_{BZ} \xi = 0.1$). It can be seen that for high values of the spin fluctuation rate the relaxation is flat and simple exponential and that there is a transition to Gaussian behavior below $\Gamma q_{BZ}^{2} \sim 10 \mu s^{-1}$. 

\section{Power law decay of spin correlations}

The mode-coupling theory of spin dynamics provides an accurate description of the long wavelength properties of spin fluctuations\cite{sw1,sw2,cuccoli,balucani}. The spin correlation function is predicted to be a homogeneous function of the variable $q^{\theta} t$ for long wavelengths and times, where $\theta$ depends on the spin model chosen. This scaling has also been predicted for a simple ferromagnet near criticality\cite{hubbardjap} for $\theta = 5/2$. Using the spin correlation function given by 
\begin{equation}
  \Lambda^{\beta \gamma} (q, t) = \frac{k_{B} T N}{g^{2} \mu_{0} \mu_{B}^{2} } \chi^{\beta \gamma} (q) f^{\beta \gamma}(q^{\theta} t).
\label{hfn}
\end{equation}
Then the Laplace transform of this is 
\begin{equation}
  \Lambda^{\beta \gamma} (q, s) = \frac{k_{B} T N}{g^{2} \mu_{0} \mu_{B}^{2} } \chi^{\beta \gamma} (q) q^{-\theta} {\cal F}^{\beta \gamma}(q^{-\theta} s),
\end{equation}
where $q^{-\theta} {\cal F}^{\beta \gamma}(q^{-\theta} s)$ is the Laplace transform of $f^{\beta \gamma}(q^{\theta} t)$. This is a general form of the spin correlation function and does not include the exponential form of the preceeding section as that is not a homogeneous function \cite{note}. It is, however, a general form of the power-law time decay of the spin correlation function that is predicted for simple spin diffusion \cite{sw2}. The Laplace transformed shape function is given by
\begin{eqnarray}
\Psi_{z} (s) &=&  \Psi_{z}^{(0)} \sum_{\beta, \gamma = x,y,z} {\cal B}^{\beta \gamma} \frac{\chi^{\beta \gamma} (0) }{\theta s^{3-(D/\theta)}} \nonumber \\  &\times& \int^{\infty}_{q_{BZ}^{-\theta} s} \frac{x^{-D/\theta} {\cal F}^{\beta \gamma}(x)}{1 + \xi^{2} \left(\frac{s}{x} \right)^{2/\theta}} dx,
\label{psiz4}
\end{eqnarray}
where $x=q^{-\theta}s$. The integral, assuming it is finite and exists, for small $s$ becomes independent of $s$. In this limit the shape function tends to
 \begin{equation}
\Psi_{z} (s) \simeq  \Psi_{z}^{(0)} \sum_{\beta, \gamma = x,y,z} {\cal B}^{\beta \gamma} \frac{I_{{\cal F}}^{\beta \gamma}}{\theta s^{3-(D/\theta)}}  ,
\label{psiz5}
\end{equation}
where the integral has been replaced by $I_{{\cal F}}^{\beta \gamma}$. Equation \ref{psiz5} has the simple inverse transform of 
 \begin{equation}
\Psi_{z} (t) \simeq  \Psi_{z}^{(0)} t^{2-(D/\theta)} \sum_{\beta, \gamma = x,y,z} {\cal B}^{\beta \gamma} \frac{I_{{\cal F}}^{\beta \gamma}}{\theta \tilde{\Gamma}\left(3 - (D/\theta)\right)}, 
\label{psiz6}
\end{equation}
where $\tilde{\Gamma}(x)$ is the Gamma function and not a relaxation rate. 

Spin diffusion\cite{sw1} gives a value of $\theta = 2$, which in three dimensions gives root exponential relaxation 
\begin{equation}
P_{Z} (t) \simeq P_{Z} (0) \exp\left\{-\lambda_{1/2} \sqrt{t}\right\},
\end{equation}
where $\lambda_{1/2}$ is given by the prefactor of $t^{2 - (D/\theta)}$ in equation \ref{psiz6} for $D=3$ and $\theta =2$. This root exponential relaxation has not been predicted before from a dynamical approach but has been observed many times \cite{blundell,jestadt,bewley}. 

For a three-dimensional ferromagnet near its critical temperature, $\theta = 5/2$ and $D=3$, this predicts that the muon will relax with the time-dependence
\begin{equation}
P_{Z} (t) \simeq  P_{Z} (0) \exp\left\{- \lambda_{4/5} t^{4/5} \right\}.
\end{equation}
It would be an interesting test if this form of the muon depolarization spectra could be observed in a model ferromagnet.   

Therefore this suggests that when stretched exponential relaxation of the muon polarization is observed, the spin correlation function is likely to be of the form of equation \ref{hfn} and possibly a power-law time decay. The stretched exponent is also directly related to the microscopic form of the spin correlation function.\\[0.05in] 

\section{Conclusions}

A method that develops the theory of spin fluctuations and their effect on the dynamic contribution to muon asymmetry has been presented. Two forms of the spin correlation function have been considered, corresponding to high temperatures in the paramagnetic regime and when the magnetic system is in a scaling regime (close to criticality). 

In the first case the spin correlation function was considered to be a simple exponential with a $q$-dependent relaxation rate. The muon depolarization spectrum was found to be an $q$-integrated Abragam function (equation \ref{psiz3}) that in the fast and slow fluctuation limits recovered the forms predicted by stochastic spin theory\cite{hayano}. The asymmetry spectra were seen to be a simple exponential when the spin fluctuation rate is large, say at high temperatures in the paramagnetic regime, and exhibited a fast drop when the shape changed to a Gaussian when the spin fluctuation rate slowed down. The transition between these two limits depends on the specifics of the spin model chosen. 

The second case considered was one where the spin correlation function is a homogeneous function of a scaling variable $q^{\theta} t$. This form was suggested by the spin diffusion and mode-coupling theories and is consistent with a power-law time-decay of the spin correlation function. In this case the asymmetry is found to be a stretched exponential with a stretched exponent of $\kappa = 2 - (D/\theta)$ where $D$ is the dimensionality and $\theta$ is a coefficient that depends on the microscopic model of the spins. Root exponential relaxation occurs in three dimensions for simple spin diffusion and $\kappa = 4/5$ for a three-dimensional ferromagnet in its scaling regime. 

Muons provide a useful alternative probe of spin fluctuations. The data can often be obtained quicker than with neutrons and for samples where neutron spectroscopy is not suitable. It is hoped that it has been demonstrated that the shape of the muon depolarization spectrum contains useful information about the type of spin correlations present and that the analysis presented will help to stimulate discussion in this area.

\begin{acknowledgements}
The author acknowledges support from the Engineering and Physical Sciences Research Council grant GR/S05267/01 and useful discussions with Professor Stephen Blundell. 
\end{acknowledgements}

\bibstyle{amsplain}

%\section*{Bibliography}
\bibliography{muon}

\begin{thebibliography}{21}
\bibitem[{not()}]{note}
\bibinfo{note}{A homogeneous function of order $n$ is one that satisfies $f(\nu
  t) = \nu^{n} f(t)$.}
\end{thebibliography}
\end{document}